\documentclass[epjCONF, onecolumn]{svjour}

\usepackage{graphics}
\usepackage[varg]{txfonts} % Times fonts
\usepackage[latin1]{inputenc}

\session-title{Assembling the puzzle of the Milky Way: Le Grand-Bornand, April 17-22, 2011}

\begin{document}

\title{Galactic substructure traced by RR Lyraes in SDSS Stripe 82}
\author{Laura Watkins\inst{1,2}\fnmsep\thanks{\email{watkins@mpia.de}}}

\institute{Max-Planck-Institut f\"{u}r Astronomie, K\"{o}nigstuhl 17, Heidelberg 69117, Germany \and Institute of Astronomy, University of Cambridge, Madingley Road, Cambridge, CB3 0HA, UK}

\abstract{
Using a sample of 407 RR Lyrae stars extracted from SDSS Stripe 82, we study the degree of substructure in the Galactic halo.  We identify overdensities associated with the known substructures of the Hercules-Aquila Cloud and the Sagittarius Stream, and find a further previously-unknown substructure, the Pisces Overdensity, at $\sim100$\,kpc from the Sun.  Together, the three substructures account for $\sim80\%$ of our RR Lyrae sample.  We also study the density distribution of RR Lyraes in the halo and find that the profile is best fit by a broken power law with an inner slope of -2.4 and a break radius of $\sim25$\,kpc, consistent with previous studies.  We conclude that the halo is predominantly made up of the debris from in-falling satellites, with little or no underlying smooth component.
}

\def\Msun{\rm M_\odot}

\maketitle

% ------------------------------------ %

\section{Introduction}
\label{sect:intro}

Simulations of the formation of our Galaxy suggest that its halo must have formed, either entirely (e.g. \cite{bullock2005}) or in part (e.g. \cite{abadi2006}), via a series of minor mergers; small subhalos caught by the potential of the Milky Way (MW) fall in and are tidally stripped as they do so, leaving streams of stars along their orbits.  These streams and the remnants of the progenitor satellites might be seen as overdensities in the halo, telling the story of the formation history of the Galaxy.  Here we study substructure using RR Lyrae stars in a $\sim300$-square-degree patch of sky known as Stripe 82.

RR Lyraes are variable stars that are commonly used as tracers of substructure (e.g. \cite{ivezic2000,vivas2004,vivas2006}); their distinctive lightcurves make them easy to identify and their well-defined absolute magnitudes allow for reliable distance estimates.  Furthermore, they are believed to be a common phase of stellar evolution, so they are ubiquitous, and they are bright, so that they can be detected out to large distances.

Stripe 82 is an equatorial strip of the SDSS footprint in the Southern hemipshere, where the observing strategy was to observe a small area many times.  The result is a data set (e.g. \cite{adelman-mccarthy2008}) with an eight-year baseline for which we have an average of 30 epochs per object (and as many as 80 epochs in the most extreme cases), making it well suited for identifying and studying variable stars.  Furthermore, there is data in five photometric bands, important for classifying and extracting stellar populations, and a subset of the data have spectra and, hence, metallicity estimates.

Bramich et al. (\cite{bramich2008}) realised the huge value of such a data set and created two catalogues in order to best exploit the data.  They are the Light-Motion-Curve Catalogue (LMCC), which contains lightcurves for all objects in the stripe, and the Higher-Level Catalogue (HLC), which contains over 200 derived quantities, e.g. mean magnitudes, mean colours and variability statistics.

Stripe 82 is of further interest because the Hercules-Aquila Cloud (\cite{belokurov2007}) is known to extend into the area covered by the stripe and the orbit of the Sagittarius Stream (e.g. \cite{belokurov2006}) also crosses the stripe.

We selected a sample of 407 RR Lyraes in Stripe 82 using the LMCC and HLC via a series of colour, period and metallicity cuts.  We then used the information in the catalogues to calculate absolute magnitudes for our sample and then distances accurate to $\sim5\%$.  Further details on the selection method can be found in \cite{watkins2009}.  Here, we present the main results from the analysis of our RR Lyrae sample.

% ------------------------------------ %

\section{Substructure in the halo}
\label{sect:substr}

\begin{figure}
	\begin{center}
        \resizebox{0.5\columnwidth}{!}{ \includegraphics{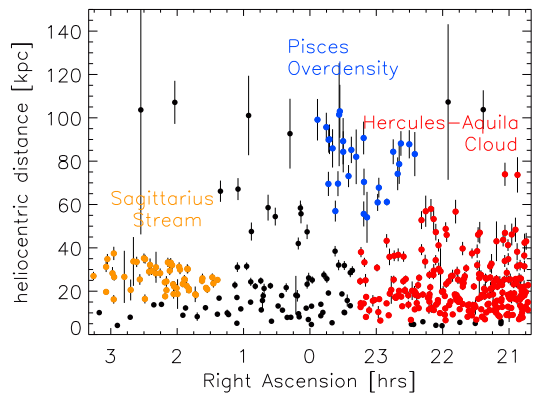} }
	    \caption{The distribution of RR Lyraes in SDSS Stripe 82 in heliocentric distance and Right Ascension.  The distribution is not uniform in Right Ascension and does not fall off smoothly with distance, suggesting that there is a lot of substructure in this part of the halo.  Three distinct substructures can be identified: the Hercules-Aquila Cloud (highlighted in red) and the Sagittarius Stream (highlighted in orange) were both previously known to exist.  The Pisces Overdensity (highlighted in blue) was identified during the course of this study.}
        \label{fig:radist}
	\end{center}
\end{figure}

The final sample of RR Lyraes is shown as function of heliocentric distance and Right Ascension (RA) in Figure \ref{fig:radist}, along with the distance errors.  The distribution is not smooth in RA, neither does the number of RR Lyraes fall off smoothly with distance; clearly the halo is far from uniform in this part of the sky.

We are able to identify three significant substructures in our distribution (for details of member selection see \cite{watkins2009}).  There is a large overdensity at high RA and small distance, highlighted in red, that has both distance and RA consistent with the Hercules-Aquila Cloud.  This substructure alone contains 237 stars, nearly 60$\%$ of our sample.  In this range of distance and RA, we would expect to find a smooth bulge component present in our sample as well as RR Lyraes belonging to the halo.  We attempted to separate the two samples using metallicities, however they proved impossible to disentangle.  We found that our metal-rich, intermediate metallicity and metal-poor samples all had the same spatial distribution (see \cite{watkins2009}, Figure 15).

At low RA and small distance, we find a second clump of RR Lyraes which have positions and distances consistent with the Sagittarius Stream where it crosses the stripe.  Our Sagittarius sample, highlighted in orange, contains 55 stars, which is a further 15$\%$ of the total sample.

The third overdensity - the Pisces Overdensity - is highlighted in blue.  It is more distant at $\sim$100 kpc and contains 28 stars, or 7$\%$ of our sample; it looks as if the structure could be bifurcated, although with so few stars, it is not possible to be certain.  This structure was also identified by Sesar et al. (\cite{sesar2007}, ``Structure J") although they did not confirm whether it was real or simply an artefact in their data.

First we ask: is it real? The error bars at these distances are large enough to be visible in the figure but not large enough to explain the presence of this substructure.  We also analysed extinction changes across the stripe but found very little variation and conclude that Pisces is indeed a real structure.

Models of the Sagittarius Stream do not predict that the stream would be detected in this part of the sky, so it is unlikely that this clump is associated.  The Magellanic Stream (MS) has been detected in this part of the sky, but is found at smaller distances.  Furthermore, the MS has only been traced in gas and is very diffuse in this region; if we were to detect the first stars in the MS then it would be more likely in a denser region.  So we are unable to associate Pisces with any known substructure.

We can make some simple comparisons with other known objects in the halo in order to learn more about the Pisces Overdensity; although, of course, any comparisons that we make can only tell us about the part of Pisces that we have detected in the stripe.  First we compare Pisces with MW dwarf spheroidal (dSph) galaxy Carina (\cite{dallora2003}), which also lies at a heliocentric distance of $\sim$100 kpc.  Assuming similar stellar populations and detection efficiencies and using simple scaling arguments, we estimate a mass of $\sim10^5 \Msun$ for Pisces in the stripe.

We can also compare with our Hercules-Aquila sample, which has the advantage of being detected via the same method, but this cloud is less well understood than Carina.  Again, we assume similar stellar populations to calculate a mass of $\sim10^4 \Msun$ for Pisces in the stripe, reassuringly similar to our previous estimate.  We also find that Pisces is 4-5 times more diffuse than Hercules-Aquila in the stripe.

Since we completed this work, there has been follow-up spectroscopy for a handful of stars in the Pisces Overdensity.  Kollmeier et al. (\cite{kollmeier2009}) obtained spectra for eight stars, of which five had very similar velocities.  The dispersion of these stars suggests that Pisces is a disrupted dSph galaxy and not a disrupted globular cluster (GC), although whether the remnant is bound or unbound is unclear.  The remaining three stars from their study were clearly offset from the main population but had similar velocities to each other, which suggests that there may indeed be a bifurcation, or that we could be looking at two different objects.  Sesar et al. (\cite{sesar2010}) obtained spectra for five stars, of which one was also in the Kollmeier et al. (\cite{kollmeier2009}) sample, bringing the total of Pisces stars with spectra to 12.  They also found evidence for two kinematic groups and concluded that Pisces is spatially extended and is likely to be unbound.

% ------------------------------------ %

\section{Density distribution of RR Lyraes in the halo}
\label{sect:density}

\begin{figure}
	\begin{center}
        \resizebox{0.45\columnwidth}{!}{ \includegraphics{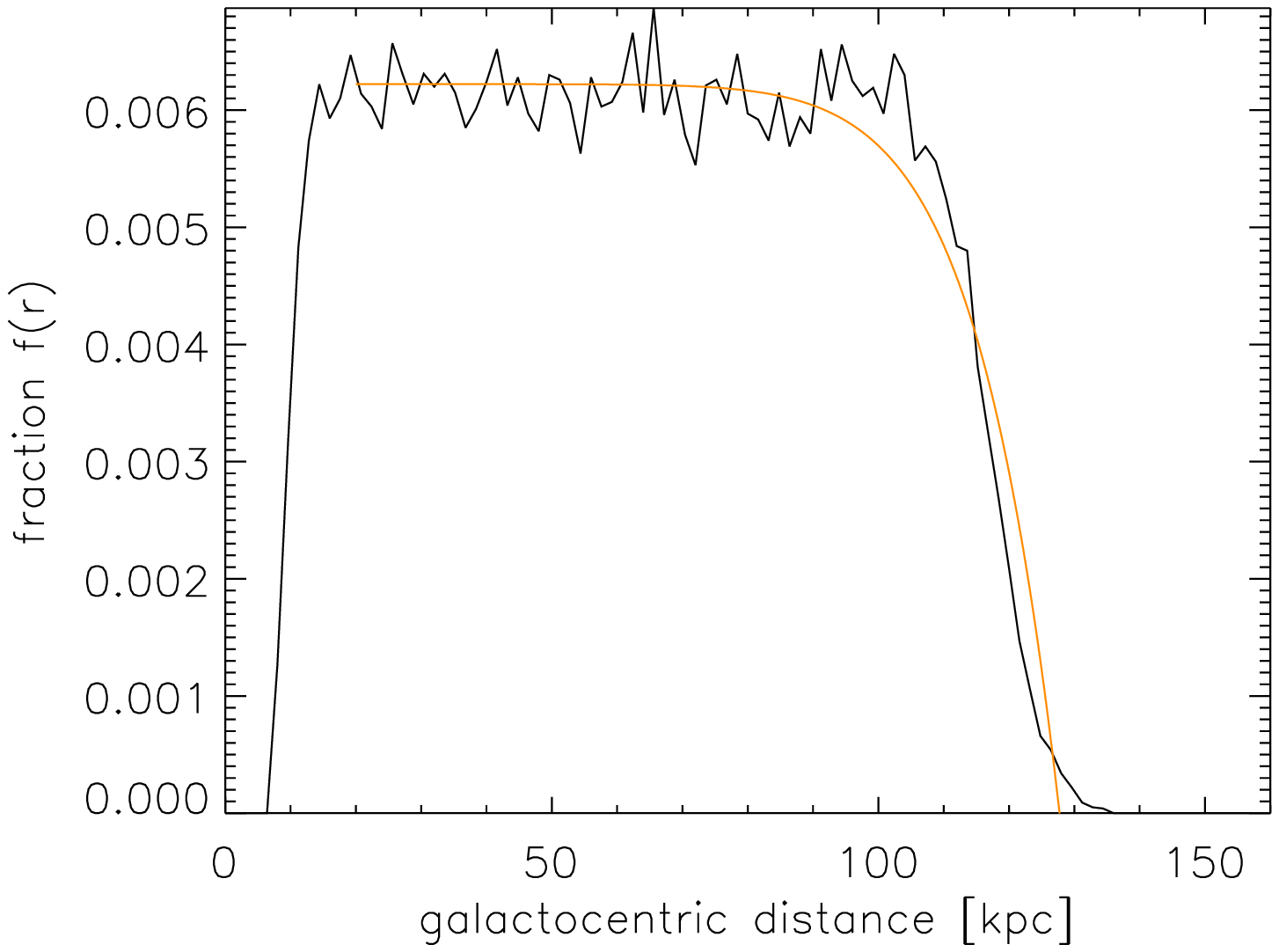} }
        \resizebox{0.45\columnwidth}{!}{ \includegraphics{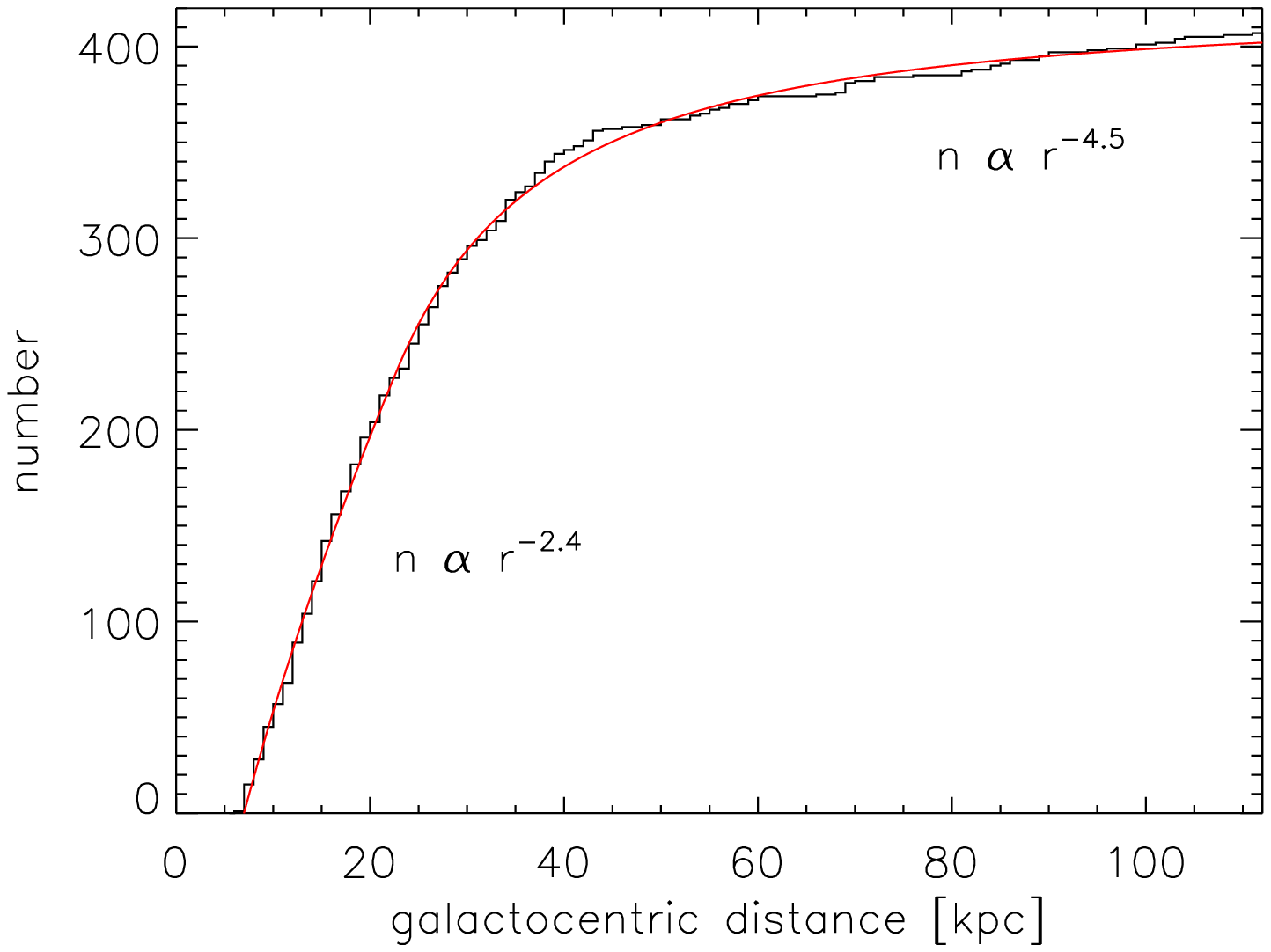} }
	    \caption{Left: the fraction of accessible volume probed by the SDSS in Stripe 82 as a function of Galactocentric distance.  It is possible to detect RR Lyraes out to 100 kpc and the brightest RR Lyraes out to 130 kpc.  Right: the cumulative number of RR Lyraes with Galactocentric distance; this distribution is best fit by a broken power law (red line).  We also see a sharp drop off beyond $\sim$40 kpc, though it is not clear whether this is a real property of the halo or due to the large amount of substructure present in Stripe 82.}
        \label{fig:density}
	\end{center}
\end{figure}

In order to study the number density of RR Lyraes in the halo, we first need to consider the fraction of accessible volume probed by the SDSS in Stripe 82 as a function of Galactocentric distance, which we show in the left panel of Figure \ref{fig:density}.  We generated RR Lyrae populations, assuming a Gaussian luminosity function with a mean and dispersion the same as that found for our RR Lyraes.  The magnitude limits were set to be the same as those defining our sample.  We see that the survey reaches out to at least 100 kpc before it starts to decline, and the brightest ($M_{\textrm{z}} \sim$ 0.1) RR Lyraes are visible all the way out to 130 kpc.

In the right panel of Figure \ref{fig:density}, we show the cumulative number of RR Lyraes with Galactocentric distance, where we have adjusted by the fraction of Galactocentric volume sampled and assumed a selection efficiency of 1.  Smooth halo models often invoke density power laws with indices of $\sim$3, but we find the best fit to the data is a double power law, which we show in red.  Such models were first proposed by Saha (\cite{saha1985}), who noticed that the density fell off more rapidly beyond 25 kpc; our estimate of the turn-over radius of 23 kpc is in excellent agreement.  And our estimate of -2.4 of the inner slope is also consistent with a study by Miceli et al. (\cite{miceli2008}) of a very large sample of RR Lyraes out to 30 kpc.

We see a sharp drop off in the number density of RR Lyraes beyond $\sim$40 kpc.  The idea of such a drop off in RR Lyrae number is certainly not new, but it has not be wholly agreed upon.  Ivezic et al. (\cite{ivezic2000}) found evidence for an ``edge" in a study of RR Lyraes found in SDSS commissioning data; although, they later applied the same method to the a larger sample that extended out to 70 kpc and found no such edge anymore.  Vivas $\&$ Zinn (\cite{vivas2006}) also conducted a similar study and found no edge out to 60 kpc.  Our data suggests that ``edge" is perhaps too strong a term for the phenomenon, as the number of RR Lyraes declines sharply rather than being completely cut off, but there is definitely a drop off and it is real.  It cannot be due to the properties of the survey - remember, we are able to see even the faintest RR Lyraes out to 100 kpc.

Of course, we saw in the previous section that there is a lot of substructure in the stripe, so its arguable how much we can really say about the density of the halo from this data.  The Hercules-Aquila Cloud alone accounts for almost 60$\%$ of our RR Lyraes, and all 3 substructures contain nearly 80$\%$ of the RR Lyraes.  Hercules-Aquila and Sagittarius dominate the stripe out to $\sim$40 kpc, so it could be their extents that explain the double-power law that we find.  Possibly there is little or no underlying smooth component in the halo and we are seeing mostly substructure.

% ------------------------------------ %

\section{Summary}
\label{sect:summary}

We extracted a set of RR Lyrae stars from the SDSS DR6 data of the region known as Stripe 82 and used them to study substructure in the halo.  We recovered two substructures that had been previously identified, and found a third that had been suggested, though not confirmed.  This Pisces Overdensity is found at $\sim$100 kpc and cannot be connected to any known substructures.  The Magellanic Stream is also found in that part of the sky but it has only been traced in gas and is found at much smaller distances.  The size of the errors and extinction changes across the stripe cannot explain it away.

We also looked at the distribution of RR Lyraes in the halo.  We found that RR Lyraes in Stripe 82 are visible out to 100 kpc, and that the brightest are visible all the way out to 130 kpc.  We also found that the density distribution is best fit by a broken power law, with a sharp fall off at $\sim$40 kpc, consistent with previous results.  However, we should bear in mind the dominance of substructure in the stripe that might naturally explain these results.  Certainly our RR Lyrae sample suggests that, if there is a smooth background halo component underneath the complexity left behind by satellite accretions, then it is only minimal.

\qquad

\noindent \emph{Acknowledgements:} I am grateful to Wyn Evans and Vasily Belokurov for their invaluable assistance with this project. I also acknowledge funding from the Science and Technology Facilities Council.

% ------------------------------------ %


\begin{thebibliography}{}

    \bibitem{bullock2005}
    Bullock, J.S. and Johnston, K.V., ApJ \textbf{635}, (2005) 931-949

    \bibitem{abadi2006}
    Abadi, M.G., Navarro, J.F. and Steinmetz, M., MNRAS \textbf{365}, (2006) 747-758

    \bibitem{ivezic2000}
    Ivezic, {\v Z}. et al., AJ \textbf{120}, (2000) 963-977

    \bibitem{vivas2004}
    Vivas, A.K. et al., AJ \textbf{127}, (2004) 1158-1175

    \bibitem{vivas2006}
    Vivas, A.K. and Zinn, R., AJ \textbf{132}, (2006) 714-728

    \bibitem{adelman-mccarthy2008}
    Adelman-McCarthy, J.K. et al., ApJS \textbf{175}, (2008) 297-313

    \bibitem{bramich2008}
    Bramich, D.M. et al., MNRAS \textbf{386}, (2008) 887-902

    \bibitem{belokurov2007}
    Belokurov, V. et al., ApJL \textbf{657}, (2007) L89-L92

    \bibitem{belokurov2006}
    Belokurov, V. et al., ApJL \textbf{642}, (2006) L137-L140

    \bibitem{watkins2009}
    Watkins, L.L. et al., MNRAS \textbf{398}, (2009) 1757-1770

    \bibitem{sesar2007}
    Sesar, B. et al., AJ \textbf{134}, (2007) 2236-2251

    \bibitem{dallora2003}
    Dall'Ora, M. et al., AJ \textbf{126}, (2003) 197-217

    \bibitem{kollmeier2009}
    Kollmeier, J.A. et al., ApJL \textbf{705}, (2009) L158-L162

    \bibitem{sesar2010}
    Sesar, B. et al., ApJ \textbf{717}, (2010) 133-139

    \bibitem{saha1985}
    Saha, A., ApJ \textbf{289}, (1985) 310-319

    \bibitem{miceli2008}
    Miceli, A. et al., ApJ \textbf{678}, (2008) 865-887

\end{thebibliography}
\end{document}